\documentclass[12pt, a4paper]{article}
\usepackage{amssymb,amsmath,epsfig}

\begin{document}

\title{Large scale and large period limits of symmetric calorons}

\author{Derek Harland\footnote{email address: d.g.harland@durham.ac.uk}
  \bigskip
  \\Deparment of Mathematical Sciences,
  \\Durham University,
  \\DH1 3LE}
\date{18th April 2007}
\maketitle

\begin{abstract}
We construct $SU(2)$ calorons, with non-trivial holonomy, instanton charge 2 and magnetic charge 0 or -1; these calorons have two constituent monopoles, with charges $(2,2)$ or $(2,1)$.  Our calorons are $U(1)$-symmetric and are constructed via the Nahm transform.  They fall into distinct families which can be classified using representation theory.  We consider large scale and large period limits of these calorons; in particular, the large scale limit may be a monopole, or a caloron with different topological charges.
\end{abstract}

\section{Introduction}

There has long been interest in calorons, that is, finite action anti-self-dual gauge fields on $\mathbb{R}^3\times S^1$.  One reason to study calorons is to gain understanding of finite temperature Yang-Mills theory.  Another reason is that calorons provide a means of relating monopoles and instantons.  Naively, when the radius of the circle $S^1$ is larger than the size of the caloron, the caloron approximates an instanton on $\mathbb{R}^4$, while when the radius of the circle is small the caloron approaches a monopole on $\mathbb{R}^3$.

Instantons, monopoles and calorons can all be classified by topological charges.  Instantons and monopoles both posess a single topological charge (at least when the gauge group is $SU(2)$), which is an integer.  Calorons, on the other hand, posess both an instanton-like charge and monopole-like charges.  Also of importance is the holonomy of the caloron, which is a path ordered integral of the gauge field around the $S^1$.  Calorons are classified by their instanton charge, magnetic charges, and the value of the holonomy at spatial infinity.  These quantities can be combined into an overall ``caloron charge'', which is not in general an integer.

An alternative description of calorons is in terms of ``constituent monopoles''; the charges and masses of the constituent monopoles are determined by the topological charges and holonomy of the caloron.  Constituent monopoles have been used previously to constuct calorons with no magnetic charge and instanton charge 1 \cite{vb98,ll}.  The constituent monopoles description made the relation between calorons and monopoles clearer: a charge 1 $SU(2)$ caloron has two constituent monopoles, and the monopole limit of the caloron arises when the location of one monopole goes to infinity.

One of the most successful tools for constructing instantons and monopoles is the ADHM-Nahm transform \cite{cg}.  This associates to each instanton or monopole a set of data (ADHM data for instantons or Nahm data for monopoles).  In general, ADHM data and Nahm data are easier to find than instantons and monopoles, so the ADHM-Nahm transform gives a simple method to find instantons and monopoles implicitly.  It is normally possible to construct instantons and monopoles numerically if their ADHM data or Nahm data is known.

There exists a Nahm transform for calorons too, which is expressed in terms of the consituent monopoles of the caloron.  Previously, this has been used to construct calorons with no magnetic charges \cite{vb98, vb02,vb03,vb04}.  To date the Nahm transform has not been applied to calorons with non-zero magnetic charge.

This article is concerned with $SU(2)$ calorons which are symmetric under an action of $U(1)$.  We will restrict attention to calorons with caloron charge between 1 and 2.  Our approach will be based on representation theory: by considering all possible representations of symmetry groups on the Nahm data, we are able to find all symmetric calorons.  We will see that when the symmetry group is $U(1)$, there are two distinct families of symmetric calorons, which are classified by representations acting on the Nahm data.

We will investigate these families, in particular their large period and large scale limits.  This will make the relations between monopoles, instantons, and calorons clearer.  The large period limit of any of our calorons is an instanton, while the large scale limit may be a monopole or a caloron with non-zero magnetic charge.  This sheds some light on the question of when a caloron has a monopole limit, and also motivates the study of calorons with non-zero magnetic charge.

An outline of the remainder of this article is as follows: in section \ref{boundary conditions} we will review the boundary conditions and topological charges of $SU(2)$ calorons.  In section \ref{cft'h ansatz} we will review the calorons of Harrington and Shepard, in particular considering their large scale limit.  In section \ref{nahm transform} we will review the Nahm transform for calorons, and this will be applied in section \ref{symmetry} to generate families of symmetric 2-calorons.  In section \ref{large scale limits} we consider the large scale limits of these calorons, and in section \ref{large period limits} we consider their large period limits.  We conclude with some comments in section \ref{conclusion}.

\section{Topological charges and boundary conditions}
\label{boundary conditions}

The topological charges and boundary conditions for $SU(n)$ calorons were first classified in \cite{gpy}, and were more recently studied in \cite{nye}.  A different approach to caloron boundary conditions, which extends to other geometries, was considered in \cite{etesi}.  Here we review the essential features of the classification of \cite{nye}, restricting attention to the case $n=2$.

Let $x^1,\,x^2,\,x^3$ be standard coordinates on $\mathbb{R}^3$ and let $x^0$ be a coordinate on $S^1$ with period $2\pi/\mu_0$.  Let $r=\sqrt{(x^1)^2+(x^2)^2+(x^3)^2}$ be a radial coordinate, and let $z^1,z^2$ denote arbitrary local coordinates on the sphere $S^2$ at fixed $r$ and $x^0$.  We let $A = A_\alpha dx^\alpha$ denote an $SU(2)$ gauge field on $\mathbb{R}^3\times S^1$ and let $D = d + A$ denote the corresponding connection, with curvature $F$.

The boundary data for a caloron will consist of
\[ ( k_0, k, \mu_0, \mu ), \]
where $k_0,k\in\mathbb{Z}$ satisfy $k_0\geq0$ and $k_0+k \geq0$, and $\mu\in[0,\mu_0/2]$.  A caloron is a gauge field whose curvature satisfies the anti-self-duality equation,
\[ \star F= -F, \]
for which the following limits hold uniformly in $x^0$, in some local gauges near $r=\infty$:
\begin{itemize}
  \item $A_0 = \mbox{diag}(i\mu,-i\mu) - \frac{1}{2r}\mbox{diag}(ik,-ik) + O(r^{-2})$
  \item $\left\| D_{z^j} A_0 \right\| = O(r^{-1})$ and $\|\partial_{0} A_{z^j}\|=O(r^{-1})$ for $j=1,2$
  \item $\left\| D_{0} A_r \right\| = O(r^{-3})$
\end{itemize}
Here we have used $\|\cdot\|$ to denote the standard norm $\|u\|^2=(1/2) \mbox{Tr}\, (u^\dagger u))$ for $u\in su(2)$.  These boundary conditions are less strict than Nye's, but are still sufficient to demonstrate that the calorons have finite action and well-defined topological charges.

The integer $k$ is called the magnetic charge of the caloron, and is analogous to the charge of a magnetic monopole.  It has the following topological interpretation: consider the two-sphere at $r=\infty$ at a fixed value of $x^0$.  When $\mu\neq0$, the eigenvectors of $A_0$ define two line bundles over this two-sphere.  It follows from the anti-self-dual equation and boundary conditions that these line bundles have Chern number $\pm k$.

The integer $k_0$ is called the instanton charge of the caloron, and, roughly speaking, it measures a winding of the gauge field over the interior of $\mathbb{R}^3\times S^1$.  $k_0$ can be defined in terms of ``framed vector bundles'' \cite{nye}, or in terms of ``loop groups'' \cite{gm88}, or it can be computed by making a singular gauge transformation \cite{gpy}.

The action of the caloron is defined to be  
\[ S = -\frac{1}{2} \int_{\mathbb{R}^3\times S^1} \mbox{Tr} \left( F_{\alpha\beta} F^{\alpha\beta} \right)  \,d^4 x, \]
and the caloron charge is determined by the integral,
\[ Q = - \frac{1}{8\pi^2} \int_{\mathbb{R}^3\times S^1} \mbox{Tr}\,( F\wedge F). \]  
A Bogoml'nyi argument shows that $S \geq 8\pi^2 Q$ for any gauge field, with equality when the gauge field is a caloron.  One can also show that
\begin{equation}
\label{charge formula}
Q = k_0 + \frac{2k\mu}{\mu_0}.
\end{equation}

Before progressing, we briefly compare this classification to others in the literature.  The classification of \cite{gpy} is similar, with the notation $(k_0,k,\mu_0,\mu)=(\nu,q_1,2\pi/\beta,\ln\lambda_1/(i\beta))$.  In reference \cite{etesi}, a caloron satisfying the ``rapid decay condition'' has $k=0$, while a caloron satisfying the ``strong holonomy condition'' has $\mu=0$.  It is proven in \cite{etesi} that a caloron satisfying either of these conditions must have integer action, in agreement with (\ref{charge formula}).

\subsection{Constituent monopoles and the rotation map}

It is usual to declare calorons which are related by gauge transformations $g(x^\alpha)$ to be equivalent.  For a given set of boundary data, the set of equivalence classes of calorons is called the moduli space.  There also exist gauge transformations $g(x^\alpha)$ which are not themselves periodic, but which preserve the periodicity of the caloron.  These are called large gauge transformations.  Large gauge transformations change the boundary data of calorons, so different sets of boundary data (and hence different moduli spaces) can be identified through large gauge transformations.

This provides one motivation to introduce constituent monopole charges and masses for calorons.  An $SU(2)$ caloron is thought of as having $2$ constituent monopoles, and the charges and masses of these monopoles behave in a simple way under large gauge transformations.  The monopole charges are
\[ (m_1,m_2) = (k_0+k_1, k_0). \]
and the monopole masses are
\[ (\nu_1,\nu_2) = (2\mu, \mu_0-2\mu). \]
For any set of caloron boundary data, there exists a large gauge transformation $\rho$, called the rotation map, which swaps the monopole masses and charges:
\[
\rho: (m_1,m_2,\nu_1,\nu_2) \mapsto (m_2,m_1,\nu_2,\nu_1).
\]
Equivalently, the action of $\rho$ on the caloron boundary data is
\[
\rho: (k_0,k,\mu_0,\mu) \mapsto (k_0+k,-k,\mu_0,\mu_0/2-\mu).
\]
Clearly, the action of $\rho$ is simpler on the monopole charges and masses.  For this reason we will choose to describe calorons in terms of the constituent monopoles; a caloron with monopole charges $m_1$ and $m_2$ will be called an $(m_1,m_2)$-caloron.  The rotation map is thought to be the only large gauge transformation which changes the boundary data.

As has already been mentioned, the definitions of topological charges, boundary data and constituent monopole charges and masses can all be extended to gauge group $SU(n)$ \cite{nye}.  An $SU(n)$ caloron has one instanton charge and $n$ magnetic charges, but the sum of the magnetic charges must be zero.  An $SU(n)$ caloron has $n$ constituent monopoles.

Mathematically, the existence of constituent monopoles in calorons follows from Garland and Murray's loop group interpretation \cite{gm88}: a caloron is a monopole whose gauge group is a loop group, and as such will have one constituent monopole for each root of this group.  Further evidence for constituent monopoles comes from the form of the Nahm transform, and more recently the constituent monopoles have been given an interpretation in terms of D-brane configurations \cite{lee&yi}.  As we shall see later, the constituent monopoles can in many cases be identified as lumps of action density.

\subsection{Review of existing work on calorons}

We will briefly review studies to date on $SU(2)$ calorons.  The earliest examples of calorons were constructed by Harrington and Shepard \cite{hs}.  These were $(N,N)$-calorons, with monopole masses $(0,\mu_0)$ (when one monopole is massless, the caloron is said to have trivial holonomy).  These will be discussed further in section \ref{cft'h ansatz}.  More $(N,N)$-calorons with trivial holonomy were given by Ward in \cite{ward}, making use of symmetry and the Nahm transform.

Calorons with non-trivial holonomy were first considered by Lee and Lu \cite{ll} and Kraan and Van Baal \cite{vb98}.  Both of these papers construct $(1,1)$-calorons with non-trivial holonomy.  Higher charge calorons with non-trivial holonomy were constructed in \cite{vb02}, for which the monopole charges were $(N,N)$, and further examples with $N=2$ were given in \cite{vb03,vb04}.  We shall encounter some of these in section \ref{symmetry}.

Monopoles give further examples of calorons \cite{gpy}.  A charge $N$ monopole consists of an $su(2)$ gauge field $A^M$ and an $su(2)$ Higgs field $\Phi$ on $\mathbb{R}^3$ solving the Bogomolnyi equation,
\[ D_i \Phi = \frac{1}{2}\epsilon_{ijk} F_{jk}, \]
with the boundary condition $\|\Phi\|\rightarrow v$ as $r\rightarrow\infty$, for some $v>0$.  We can make a caloron with any period just by setting $A=A^M + \Phi dx^0$, which automatically solves the anti-self-dual equation, and this will have boundary data
\[ (k_0,k,\mu_0,\mu) = (0,N,\mu_0,v). \]
If $v\leq\mu_0/2$, all of the caloron boundary conditions are satisfied.  This is an $(N,0)$-caloron; in fact, all $(N,0)$-calorons can be obtained in this way.  If $v>\mu_0/2$, we must make a gauge transformation, $g = \exp(i m\Phi x^0/2)$ for $m\in\mathbb{Z}$, which has the effect $k_0\mapsto k_0+m$, $\mu\mapsto\mu-m\mu_0/2$.  Choosing $m=\lfloor 2v/\mu_0 \rfloor$ gives a caloron in the correct gauge, and we see that 
\[
(m_1,m_2,\nu_1,\nu_2) = \left( N+\left\lfloor \frac{2v}{\mu_0}\right\rfloor,\,\left\lfloor \frac{2v}{\mu_0} \right\rfloor,\, 2v-\mu_0\left\lfloor \frac{2v}{\mu_0} \right\rfloor,\,\mu_0\left( \left\lfloor \frac{2v}{\mu_0} \right\rfloor +1 \right) - 2v \right).
\]

Less trivial examples of calorons with $m_1\neq m_2$ have been constructed by Chakrabarti \cite{chakrabarti87} using a $U(1)$-symmetric ansatz.

\section{The Corrigan-Fairlie-t'Hooft ansatz}
\label{cft'h ansatz}

In this section we consider the calorons of Harrington and Shepard \cite{hs} in more detail; in particular, we study their large scale limit.  This will give some new examples of calorons with non-zero magnetic charge $k$.  Harrington and Shepard's construction was based on the Corrigan-Fairlie-t'Hooft ansatz for instantons on $\mathbb{R}^4$ \cite{cf},
\[ A_\mu = \frac{i}{2}\eta_{\mu\nu} \partial_\nu \ln\phi, \]
where $\rho$ is a real function, $\eta_{\mu\nu}=\eta_{\mu\nu}^a\sigma^a$, $\eta_{\mu\nu}^a = \epsilon_{0a\mu\nu} + \delta_{\mu a} \delta_{\nu0} - \delta_{\mu0} \delta_{\nu a}$ is the self-dual t'Hooft tensor, and $\sigma^a$ are the Pauli sigma matrices (satisfying $\sigma^a\sigma^b=\delta^{ab}+i\epsilon_{abc}\sigma^c$).  It can be shown that the anti-self-dual equation is solved when $\phi$ satisfies the Laplace equation, $\partial_\mu\partial^\mu \phi = 0$.  The Corrigan-Fairlie-t'Hooft $N$-instantons are those for which
\[ \phi = 1 + \sum_{j=1}^N \frac{\lambda_j^2}{|x-a_j|^2} \]
for some $\lambda_j\in\mathbb{R}$ and distinct $a_j\in\mathbb{R}^4$.

Harrington-Shepard calorons are formed by arranging the $a_j$ in an infinite periodic chain.  If one takes
\[ a_j = (2\pi j/\mu_0,0,0,0) ,\, \lambda_j = \lambda/\sqrt{\mu_0} \]
then $\phi$ takes the form,
\begin{equation}
\label{hs 1-caloron}
\phi = 1 + \frac{\lambda^2}{\mu_0} \sum_{j=-\infty}^\infty \frac{1}{|x-a_j|^2} = 1 + \frac{\lambda^2}{2r} \frac{\sinh(\mu_0r)} { \cosh(\mu_0 r)-\cos(\mu_0 x^0) }.
\end{equation}
The resulting caloron has monopole charges $(1,1)$ and monopole masses $(0,\mu_0)$.  More generally, consider the function
\begin{equation}
\label{hs N-caloron}
\phi = 1 + \sum_{j=1}^N \frac{\lambda_j^2}{2r_j} \frac{\sinh(\mu_0r_j)} { \cosh(\mu_0 r_j)-\cos(\mu_0 (x^0-t_j)) } 
\end{equation}
where $\lambda_j>0$, $r_j=|\mathbf{x}-\mathbf{a}_j|$ for $\mathbf{a}_j\in\mathbb{R}^3$, $t_j\in S^1$ and the points $(\mathbf{a}_j,t_j)$ are distinct.  This generates a caloron with monopole charges $(N,N)$ and monopole masses $(0,\mu_0)$, and will be called a Harrington-Shepard $(N,N)$-caloron.  A Harrington-Shepard caloron is normally thought of as consisting of $N$ $(1,1)$-calorons, having locations $(t_j,\mathbf{a_j})$ and scales $\lambda_j$.  These can be seen as distinct lumps of action density when the scales are small compared with the period, $2\pi/\mu_0$.  When the period becomes infinitely large, one recovers the instantons of Corrigan, Fairlie and t'Hooft.

It is interesting to consider the opposite limit of Harrington-Shepard calorons, that is, the limit where the scales are large compared with the period.  Letting all of the $\lambda_j\rightarrow\infty$ in equation (\ref{hs N-caloron}) is equivalent to removing the constant term, since the gauge field is unaffected by constant rescalings of $\phi$.  We are left with
\begin{equation}
\label{jnr N-caloron}
\phi = \sum_{j=1}^N \frac{\lambda_j^2}{2r_j} \frac{\sinh(\mu_0r_j)} { \cosh(\mu_0 r_j)-\cos(\mu_0 (x^0-t_j)) }.
\end{equation}
This function generates genuine calorons, which will be called JNR calorons, owing to the resemblance between (\ref{jnr N-caloron}) and the Jackiw-Nohl-Rebbi ansatz for instantons \cite{jnr}.  The $x^0$-component of a JNR caloron gauge field at infinity has the form
\[ A_0 = \frac{x^j}{r}\frac{\sigma^j}{2i} \left(-\frac{1}{r} + O(r^{-2})\right) \]
for large $r$.  Therefore these calorons have $k=-1$ and $\mu=0$.

One can evaluate $k_0$ for these calorons by computing the caloron charge $Q$ and using the formula (\ref{charge formula}).  $Q$ can be evaluated using a method due to Jackiw et al \cite{jnr}.  Since the gauge field is anti-self-dual, $Q=S/(8\pi^2)$.  In the CFt'H ansatz the action density is given by $\mbox{Tr}(F_{\alpha\beta} F^{\alpha\beta}) = \Box\Box\ln\phi$ \cite{jnr} (except at the points where $\phi$ is singular).  Therefore the charge is found by evaluating
\begin{equation}
\label{jnr charge}
Q = - \frac{1}{16\pi^2} \int_{\mathbb{R}^3\times S^1} \Box\Box\ln\phi\, d^4x,
\end{equation}
while ignoring singularities.  To remove the singularities, we replace $\phi$ by
\[ \phi' = \sum_{j=1}^N \frac{\lambda_j^2}{2r_j} \sinh(\mu_0 r_j)\prod_{i\neq j} (\cosh(\mu_0 r_i)-\cos(\mu_0 (x^0-t_i))). \]
This does not change the value of the integral, since $\Box\Box\ln(\cosh(\mu_0r_i)-\cos(\mu_0(x^0-t_i))=0$ away from the singularities of $\phi$.

When $r$ is large, $\phi'$ has the approximate form
\[ \phi' \approx \left(2^{-N-1}\sum_{j=1}^N \lambda_j^2\right) \frac{\exp(N\mu_0r)}{r}. \]
Since $\Box\ln(r\phi'/\exp(N\mu_0 r))=O(r^{-2})$ we have,
\begin{eqnarray*}
\Box\ln\phi' &=& \left( \partial_r^2 + \frac{2}{r}\partial_r \right) (N\mu_0r-\ln r) + O(r^{-2}) \\
&=& \frac{2N\mu_0}{r} + O(r^{-2}).
\end{eqnarray*}

This large $r$ expansion is used to evaluate (\ref{jnr charge}).  Since $\Box\Box\ln\phi'$ has no singularities, we may apply Stokes' theorem, obtaining
\begin{eqnarray*}
Q &=& -\frac{1}{16\pi^2} \lim_{R\rightarrow\infty} \int_{S^2\times S^1} \partial_r \Box \ln \phi'|_{r=R} R^2 d\Omega_2 dx^0 \\
&=& -\frac{1}{16\pi^2} \lim_{R\rightarrow\infty} \int_{S^2\times S^1} \left(-\frac{2N\mu_0}{R^2}+O(R^{-3})\right) R^2 d\Omega_2 dx^0 \\
&=& \frac{1}{16\pi^2} \lim_{R\rightarrow\infty} \left( 2N\mu_0 (4\pi)(2\pi/\mu_0) + O(R^{-1}) \right) \\
&=& N.
\end{eqnarray*}
It follows from (\ref{charge formula}) that $k_0=N$.  So we see that the JNR caloron (\ref{jnr N-caloron}) is an $(N-1,N)$-caloron, with constituent monopole masses $(0,\mu_0)$.

Things are particularly simple when $N=1$: the JNR $(0,1)$-caloron generated by 
\[ \phi = \frac{1}{2r} \frac{\sinh(\mu_0r)} { \cosh(\mu_0 r)-\cos(\mu_0 x^0) } \]
has only one constituent monopole, so we expect it to be related to a 1-monopole by a large gauge transformation.  This result was in fact obtained long ago by Rossi \cite{rossi}; Rossi's gauge transformation was an explicit example of a rotation map.

There is a final limit which we will consider.  In equation (\ref{jnr N-caloron}), one can send one of the locations $\mathbf{a}_j$ to infinity.  Consider the limit $|\mathbf{a}_N|\rightarrow\infty$, $\lambda_N\rightarrow\infty$ such that $\lambda_N^2/(2|\mathbf{a}_N|)\rightarrow1$.  Then we are left with a function,
\[ \phi = 1 + \sum_{j=1}^{N-1} \frac{\lambda_j^2}{2r_j} \frac{\sinh(\mu_0r_j)} { \cosh(\mu_0 r_j)-\cos(\mu_0 (x^0-t_j)) }. \]
The resulting caloron is of course a Harrington-Shepard $(N-1,N-1)$-caloron.  So, to summarise this discussion, we are able to obtain JNR calorons and Harrington-Shepard calorons as limits of each other, and these limiting processes always reduce the number of constituent monopoles.

\section{The Nahm transform for calorons}
\label{nahm transform}

The Nahm transform is a bijection between calorons and sets of ``Nahm data''.  The Nahm transform therefore provides a means of constructing calorons implicitly, and it usually possible to determine interesting properties of calorons from their Nahm data.  In principle the Nahm transform can be formulated for calorons with gauge group $SU(n)$, $Sp(n)$, or $SO(n)$ by adapting the well-studied monopole Nahm transforms \cite{hurt&murr89}.  In practise only the $SU(n)$ Nahm transform has been rigorously analysed \cite{nye}, and the $SO(n)$ Nahm transform for calorons has not even been written down.  We are interested only in gauge group $SU(2)$, and since $SU(2)$ and $Sp(1)$ are isomorphic, there are in fact two Nahm transforms applicable to our case, coming from the $SU(n)$ Nahm transforms and the $Sp(n)$ Nahm transforms.  We will choose to use the $Sp(1)$ version, which was also used by Van Baal et. al. \cite{vb98,vb02,vb03,vb04}.

In this section we will describe the Nahm data and Nahm transform for $(N,N)$-calorons and $(N,N-1)$-calorons.  We will consider the gauge freedom of the Nahm data in detail for the case $N=2$.

\subsection{The Nahm data}

The Nahm data is formulated in terms of the constituent monopoles of the caloron.  We begin with a circle $\mathbb{R}/\mu_0\mathbb{Z}$, which is divided into two intervals $I_1$, $I_2$ of length $\nu_1$ and $\nu_2$.  We let $s$ be a coordinate on the circle and take $I_1=(-\mu,\mu)$ and $I_2=(\mu,\mu_0-\mu)$.  There are two parts to the Nahm data, associated with the intervals $I_p$ and their endpoints $\pm\mu$.

The first part of the Nahm data consists of two sets of four matrix-valued functions, $T_p^\alpha(s)$, where $p=1,2$ and $\alpha=0,1,2,3$.  The Hermitian matrices $T_p^\alpha(s)$ are only defined for $s\in I_p$, and have dimension $m_p\times m_p$.  They must satisfy the reality condition,
\begin{equation}
\label{reality condition}
T_p^\alpha(-s) = T_p^\alpha(s)^t,
\end{equation}
where $^t$ denotes matrix transpose, and they must satisfy the Nahm equation,
\begin{equation}
\label{nahm equation}
\frac{d}{ds} T_p^j - i[T_p^0,T_p^j] - \frac{i}{2} \epsilon_{jkl} [T_p^k,T_p^l] = 0
\end{equation}
for $j=1,2,3$.

The second part of the Nahm data depends on the difference $m_1-m_2$.  We will only give the details for the two cases which will be relevant to our purposes.  The first case is when $m_1-m_2=0$; we will write $(m_1,m_2)=(N,N)$.  Then the second part of the Nahm data consists of an $N$ row vector $W$ of quaternions and a purely imaginary unit quaternion $\tau$ (by ``unit quaternion'', we mean $\tau^\dagger\tau=1$).  The Nahm data are required to satisfy the matching conditions,
\begin{eqnarray}
\label{matching condition 1}
T_2^j(\mu) - T_1^j(\mu) &=& i\Re (e_j W^\dagger P_1 W) \\
\label{matching condition 2}
T_1^j(-\mu) - T_2^j(\mu_0-\mu) &=& i\Re (e_j W^\dagger P_2 W)
\end{eqnarray}
for $j=1,2,3$, with $P_1:=(e_0+i\tau)/2$ and $P_2:=(e_0-i\tau)/2$ elements of $\mathbb{C} \otimes_\mathbb{R} \mathbb{H}$.  Here we have used $e_\alpha$ to denote the quaternions, satisfying $e_i e_j=-\delta_{ij}e_0 + \epsilon_{ijk}e_k$, and $\Re$ refers to the real quaternionic part.  In general $^\dagger$ will act on complex-quaternionic matrices by transposing and taking quaternion and complex conjugates; so in the preceding formulae, $W^\dagger$ just denotes the quaternionic-conjugate transpose of $W$.  This Nahm transform matches \cite{vb02} under the identifications $s=2\pi z$, $T_p^\alpha=(i/2\pi)\hat{A}_\alpha$, $W=\zeta$.

The second case we will consider is $m_1-m_2=1$; we will write $(m_1,m_2)=(N,N-1)$.  In this case the second part of the Nahm data is an $N\times (N-1)$ complex matrix $X$ satisfying $X^\dagger X = I_{N-1}$.  The Nahm data must satisfy the matching conditions
\begin{eqnarray}
\label{matching condition 3}
X^\dagger T_1^j(\mu)X &=& T_2^j(\mu) \\
\label{matching condition 4}
X^t T_1^j(-\mu)\bar{X} &=& T_2^j(\mu_0-\mu)
\end{eqnarray}
for $j=1,2,3$, where $\bar{}$ denotes complex conjugation.

\subsection{Nahm transform}

Now we describe the implementation of the Nahm transform, in other words, how to get a caloron from a set of Nahm data.  Let $U_p(s,x^\alpha)$, $p=1,2$, be two $m_p$-component column-vectors of complex quaternions, defined for $s\in I_p$ and $(x^\alpha)\in\mathbb{R}^3\times S^1$.  Let $x=x^\alpha e_\alpha$ and let $T_p=T_p^\alpha e_\alpha$.  The vectors $U_p$ must satisfy the differential equation,
\begin{equation}
\label{dirac equation}
\frac{d}{ds} U_p(s,x^\alpha) = i(T_p(s)+x) U_p(s,x^\alpha).
\end{equation}

In the case $(m_1,m_2)=(N,N)$ we must also define a single quaternion $V(x^\alpha)$, and $U_p$ and $V$ must satisfy the matching condition,
\begin{eqnarray}
\label{dirac matching 1}
U_2(\mu,x^\alpha) - U_1(\mu,x^\alpha) &=& iW^\dagger P_1 V(x^\alpha) \\
\label{dirac matching 2}
U_1(-\mu,x^\alpha) - U_2(\mu_0-\mu,x^\alpha) &=& iW^\dagger P_2 V(x^\alpha).
\end{eqnarray}
$U_p$ and $V$ must also satisfy the normalisation condition,
\begin{equation}
\label{normalisation}
\sum_{p=1}^2 \int_{I_p} U_p^\dagger U_p ds + V^\dagger V = e_0.
\end{equation}
Then the caloron gauge field is determined by
\begin{equation}
\label{gauge field}
A_\alpha = \sum_{p=1}^2 \int_{I_p} U_p^\dagger \frac{\partial U_p}{\partial x^\alpha} ds + V^\dagger \frac{\partial V}{\partial x^\alpha}.
\end{equation}

In the case where $(m_1,m_2)=(N,N-1)$, there is no need to define $V$.  The matching conditions (\ref{dirac matching 1}), (\ref{dirac matching 2}) are replaced by,
\begin{eqnarray}
\label{dirac matching 3}
X^\dagger U_1(\mu,x^\alpha) &=& U_2(\mu,x^\alpha) \\
\label{dirac matching 4}
X^t U_1(-\mu,x^\alpha) &=& U_2(\mu_0-\mu,x^\alpha).
\end{eqnarray}
The normalisation condition and the equation for the caloron gauge field are the same as in the $(N,N)$ case, except that terms involving $V$ are omitted.

\subsection{Gauge transformations: $(2,2)$ case}

There exist gauge transformations which act on the Nahm data, and Nahm data related by gauge transformations give rise to the same caloron.  Here we describe the action of gauge transformations in the case $(m_1,m_2)=(2,2)$, and in the next subsection we will consider $(m_1,m_2)=(2,1)$.  These are the cases for which we will solve the Nahm equations.

There are two kinds of gauge transformation present for $(2,2)$ Nahm data.  The first is a $U(2)$ gauge transformation, which is defined by a $U(2)$-valued function $g(s)$.  This satisfies a reality condition, $g(-s) = \bar{g}(s)$.  $g$ acts in the following way:
\begin{eqnarray*}
T_p^j(s) & \mapsto & g(s) T_p^j(s) g^{-1}(s) \\
T_p^0(s) & \mapsto & g(s) T_p^0(s) g^{-1}(s) -i \frac{dg}{ds}(s) g^{-1}(s) \\
W & \mapsto & P_1 W g^{-1}(\mu) + P_2 W g^{-1}(-\mu) \\
U_p(s) & \mapsto & g(s) U_p(s) \\
V & \mapsto & V.
\end{eqnarray*}

The second kind of gauge transformation is a quaternion gauge transformation, defined by a unit quaternion $h$.  The action of $h$ on the Nahm data is
\begin{eqnarray*}
V &\mapsto& hV \\
W &\mapsto& hW \\
\tau &\mapsto& h\tau h^\dagger,
\end{eqnarray*}
leaving $T_p^\alpha$ and $U_p$ fixed.

It is straightforward to check that these gauge transformations map sets of Nahm data to sets of Nahm data, and leave the caloron gauge field unaffected.  Given a set of Nahm data, it is always possible to make a $U(2)$ gauge transformation and a translation in $x^0$ so that
\[
T_p^0(s) = \frac{\xi}{\mu_0} ( \sigma^3\cos\theta + \sigma^1 \sin\theta ),
\]
with constants $\xi\in[0,\pi]$ and $\theta\in[0,2\pi)$, and where $\sigma^j$ are the Pauli sigma matrices.  By gauge rotation, one can then fix $\theta=0$.

\subsection{Gauge transformations: $(2,1)$ case}

We will divide the $(2,1)$ Nahm data gauge transformations into two types.  There are $U(2)$ gauge transformations $g_1(s)$ defined for $s\in I_1$, and $U(1)$ gauge transformations $g_2(s)$ defined for $s\in I_2$.  These must satisfy the reality condition $g_p(-s)=\bar{g}_p(s)$.  Their action is as follows:
\begin{eqnarray*}
T_p^j(s) & \mapsto & g_p(s) T_p^j(s) g_p^{-1}(s) \\
T_p^0(s) & \mapsto & g_p(s) T_p^0(s) g_p^{-1}(s) -i \frac{dg_p}{ds}(s) g_p^{-1}(s) \\
X & \mapsto & g_1(\mu) X g_2(\mu)^{-1} \\
U_p(s) & \mapsto & g_p(s) U_p(s).
\end{eqnarray*}

Using gauge transformations and $x^0$ translations, it is always possible to make $X=(\sin\beta/2,\,-\cos\beta/2)^t$ and fix $T_2^0=0$ and $T_1^0=(\xi/\nu_1)\sigma^3$ for some constants $\xi\in[0,\pi]$ and $\beta\in[0,2\pi)$.

\section{Calorons with $U(1)$ symmetry}
\label{symmetry}

There is a natural action of $SU(2)$ on gauge fields, where the group acts by rotation on $\mathbb{R}^3$ and by the adjoint representation on the Lie algebra.  There exists a large family of instantons on $\mathbb{R}^4$ which are invariant under this action \cite{witten}, and many monopoles have been found invariant under the action of $SU(2)$ and its subgoups.  Therefore it is reasonable to look for calorons invariant under the action of $SU(2)$ and its subgroups.  In this section, we will show how to obtain Nahm data for calorons invariant under the action of a subgroup $G$ of $SU(2)$, and we will give explicit Nahm data for $(2,2)$- and $(2,1)$-calorons which are invariant under the action of $U(1)$.

\subsection{Action of $SU(2)$ on calorons and Nahm data}

The action of $SU(2)$ on calorons is as follows.  Let $R$ be an element of $SU(2)$, and let $R_2$ denote it's image under the fundamental representation.  Let $R_3$ denote the image of $R$ in the irreducible 3-dimensional representation of $SU(2)$, so that the entries of $R_3$ are real.  If $\mathbf{x}=(x^1,x^2,x^3)$, we can write $(R_3\mathbf{x})_j=R_3^{jk}x^k$.  We have $R_2 \sigma^j R_2^{-1} = R_3^{kj}\sigma^k$.  Then the action of $R$ on a caloron gauge field is
\begin{eqnarray*}
A_j(x^0,\mathbf{x}) & \mapsto & R_3^{jk} R_2 A_k(x^0,R_3^{-1}\mathbf{x}) R_2^{-1} \\
A_0(x^0,\mathbf{x}) & \mapsto & R_2 A_0(x^0,R_3^{-1}\mathbf{x}) R_2^{-1}.
\end{eqnarray*}
A caloron is said to be $G$-symmetric for some subgroup $G\subseteq SU(2)$ if it is invariant under the action of $R$ for all $R\in G$.

If a caloron is $G$-symmetric, we expect its Nahm data to be invariant under some action of $G$.  Consider the action,
\begin{eqnarray}
\label{Tj sym}
T_p^j &\mapsto& R_3^{jk} R_N T_p^{k} R_N^{-1} \\
\label{T0 sym}
T_p^0 &\mapsto& R_N T_p^0 R_N^{-1} \\
\label{W sym}
W^\dagger &\mapsto& R_N (R_2 W^\dagger R_{2'}^{-1}) \\
\label{tau sym}
\tau &\mapsto& R_{2'} \tau R_{2'}^{-1},
\end{eqnarray}
where $R\mapsto R_N$ is a real $N$-dimensional orthogonal representation of $G$ and $R\mapsto R_{2'}$ is a 2-dimensional unitary representation of $G$.  In the above expressions, $R_2$ and $R_{2'}$ are acting on quaternions, which is made possible using the standard representation of the quaternions, $e_0=I_2$ and $e_j=-i\sigma^j$.  If a set of Nahm data is invariant under this action for some choice of representations $R_N$, $R_{2'}$, and for all $R\in G$, the corresponding caloron will be $G$-symmetric.

Similar constraints can be derived for the Nahm data of an $(N,N-1)$ caloron.  In the case $N=2$, a caloron will be $G$-symmetric if its Nahm data is invariant under the action
\begin{eqnarray}
\label{T1j sym}
T_1^j &\mapsto& R_3^{jk} R_{N} T_1^{k} R_N^{-1} \\
\label{T10 sym}
T_1^0 &\mapsto& R_N T_1^0 R_N^{-1} \\
\label{T2j sym}
T_2^j &\mapsto& R_3^{jk} T_2^{k}.
\end{eqnarray}
for all $R\in G$, where $R_N$ is the image of $R$ under some real 2-dimensional orthogonal representation of $G$.

So $G$-symmetric calorons can be found by constructing Nahm data satisfying the above symmetry conditions.  To find this Nahm data, we first must choose the representations $R_N$ and $R_{2'}$ so that the symmetry conditions have non-trivial solutions.  In terms of representation theory, we must choose $R_N$ and $R_{2'}$ so that the dimensions of the trivial subrepresentations of the representations of $G$ acting on the Nahm data are maximised.  For each such choice of $R_N$ and $R_{2'}$, we attempt to solve the Nahm equations and matching conditions with $G$-invariant Nahm data.

\subsection{$U(1)$-symmetric $(2,2)$-calorons}

We will look for $(2,2)$-calorons invariant under the action of the $U(1)$ subgroup of $SU(2)$ generated by $\sigma^2$.  Let $P_0$ denote the trivial representation of $U(1)$, and let $Q_k$ denote the real 2-dimensional representation,
\[ Q_k : \exp(i\theta \sigma^2) \mapsto \left( \begin{array}{cc} \cos k\theta & \sin k\theta \\ -\sin k\theta & \cos k\theta \end{array} \right) \]
for $k\in\mathbb{Z}$.  Note that $Q_0=2P_0$ and $Q_{-k}$ is equivalent to $Q_k$.  We will need to consider tensor products of such representations; we have
\[ Q_k \otimes Q_l = Q_{k+l} \oplus Q_{k-l}. \]

Let us consider the representations in (\ref{Tj sym})-(\ref{tau sym}).  We have $(R\mapsto R_2)=Q_1$ and $(R\mapsto R_3)=Q_2\oplus P_0$, where $P_0$ acts on $x^2$ and $Q_2$ acts on the subspace spanned by $x^1$ and $x^3$.  The representation $R_N$ is 2-dimensional and real, so it must be of the form $R_N=Q_k(R)$ for some $k\in\mathbb{Z}$.  We can make quaternion gauge transformations so that $R_{2'}=Q_l(R)$ for some $l\in\mathbb{Z}$.

Then the representation acting on the $2\times2$ Hermitian matrix $T_p^0$ in equation (\ref{T0 sym}) is
\[ Q_{2k}\oplus 2P_0, \]
where $Q_{2k}$ acts on the subspace spanned by $\sigma^1$ and $\sigma^3$, and $2P_0$ acts on the subspace spanned by $\sigma^2$ and the identity matrix $\mathbf{1}_2$.  The representation acting on $T_p^j$ in equation (\ref{Tj sym}) is a tensor product of this representation with the representation $(R\mapsto R_3)$ acting on the index $j$.  So $T_p^j$ is acted on by  
\[ (Q_2\oplus P_0) \otimes (Q_{2k}\oplus 2P_0) = Q_{2+2k} \oplus Q_{2-2k} \oplus 2Q_2 \oplus Q_{2k} \oplus 2P_0. \]

In equation (\ref{W sym}) there is a representation acting quaternions, $q \mapsto Q_1(R)q Q_l(R)^{-1}$.  One can show that this representation is equal to $Q_{1+l}\oplus Q_{1-l}$, where $Q_{1+l}$ acts on the subspace spanned by $e_1$ and $e_3$ and $Q_{1-l}$ acts on the subspace spanned by $e_0$ and $e_2$.  So the representation acting on $W$ in (\ref{W sym}) is
\[ Q_k \otimes (Q_{1+l} \oplus Q_{1-l}) = Q_{k+l+1} \oplus Q_{k-l-1} \oplus Q_{k+l-1} \oplus Q_{k-l+1}. \]

We see that the trivial subrepresentations of these three representations are largest when $(k,l)$ are equal to $(0,\pm1)$, $(\pm 1,0)$, or $(\pm 1 , \pm 2)$.  As we shall see below, the Nahm equations can be solved in two of these three cases.

\subsubsection{$R_N$ trivial}
\label{RN trivial}

First we consider the case $(k,l)=(0,\pm1)$, so that the representation $R\mapsto R_N$ is trivial.  The invariant Nahm data must take the following form:
\begin{eqnarray*}
T_p^0 &=& (\xi/\mu_0) \sigma^3 \\
T_p^j &=& 0  \\
T_1^2(s) &=& \exp(iT_1^0 s) Y_1 \exp(-iT_1^0s) \\
T_2^2(s) &=& \exp(iT_2^0 (s-\mu_0/2)) Y_2 \exp(-i T_2^0 (s-\mu_0/2)) \\
W &=& \begin{cases} \lambda ( \cos(\beta/2)e_0,\, \sin(\beta/2) \exp(\alpha e_2) ) & l=1 \\
                    \lambda e_1 ( \cos(\beta/2)e_0,\, \sin(\beta/2)\exp(\alpha e_2) ) & l=-1
      \end{cases} \\
\tau &=& e_2
\end{eqnarray*}
for $p=1,2$ and $j=1,3$.  Here $\lambda >0$, $\beta\in[0,2\pi)$ and $\alpha\in[0,2\pi)$ are constants, and we have used gauge transformations and coordinate translations to fix the form of $T_p^0$ in terms of the constant $\xi\in[0,\pi]$.  $Y_1$ and $Y_2$ are Hermitian matrices, which are constant as a consequence of the Nahm equation (\ref{nahm equation}).  We will write
\[ Y_p = Y_p^0 \mathbf{1}_2 + Y_p^j \sigma^j, \]
and the reality condition (\ref{reality condition}) implies that $Y_p^2=0$.

This Nahm data now solves the Nahm equation and the reality condition, so it only remains to consider the matching conditions (\ref{matching condition 1}), (\ref{matching condition 2}).  After some rearrangement these take the form,
\begin{eqnarray}
\nonumber Y_2^0 - Y_1^0 &=& \pm \lambda^2/4 \\
\label{explicit matching condition}
Y_2^3 - Y_1^3 &=& \pm (\lambda^2/4) \cos\beta \\
\nonumber Y_2^1\cos(\xi\nu_2/\mu_0) - Y_1^1 \cos(\xi\nu_1/\mu_0) &=& \pm (\lambda^2/4) \sin\beta\cos\alpha \\
\nonumber Y_2^1\sin(\xi\nu_2/\mu_0) + Y_1^1\sin(\xi\nu_1/\mu_0) &=& (\lambda^2/4) \sin\beta\sin\alpha,
\end{eqnarray}
where $\pm$ corresponds to the choice $l=\pm1$.  In the case where $\xi\neq0$, the solution of the matching conditions is,
\begin{eqnarray}
\nonumber Y_1^0 &=& 0 \\
\nonumber Y_2^0 &=& \pm \frac{\lambda^2}{4} \\
\nonumber Y_1^3 &=& \chi \\
\label{xi neq 0}
Y_2^3 &=& \chi \pm \frac{\lambda^2}{4} \cos\beta \\
\nonumber Y_1^1  &=& \frac{\lambda^2}{4} \frac{ \sin\beta \sin(\alpha \mp \xi\nu_2/\mu_0) }{ \sin\xi }. \\
\nonumber Y_2^1  &=& \frac{\lambda^2}{4} \frac{ \sin\beta \sin(\alpha \pm \xi\nu_1/\mu_0) }{ \sin\xi }
\end{eqnarray}
Here $\chi$ is a real number and we have fixed $Y_1^0=0$ by making an $x^2$-translation.  We see that this family of calorons depends on five parameters: $\lambda$, $\chi$, $\alpha$, $\beta$ and $\xi$.  In the case where $\xi=0$, the solution of the matching conditions can be parametrised in the following way:
\begin{eqnarray}
\nonumber Y_1^0 &=& 0 \\
\nonumber Y_2^0 &=& \pm \frac{\lambda^2}{4} \\
\nonumber Y_1^3 &=& \chi \\
\label{xi=0}
Y_2^3 &=& \chi \pm \frac{\lambda^2}{4} \cos\beta \\
\nonumber Y_1^1  &=& 0 \\
\nonumber Y_2^1  &=& \pm \frac{\lambda^2}{4} \sin\beta,
\end{eqnarray}
with $\alpha=0$.  Here the gauge has been chosen so that $Y_1^1=0$.  This corresponds to a three-parameter sub-family of the five-parameter family.

The expressions (\ref{xi neq 0}), (\ref{xi=0}) remain valid when one of the monopoles is massless, that is, when $\nu_2=0$.  In particular, when $\nu_2=0$ and $\alpha=0$, the caloron obtained via the Nahm transform is a Harrington-Shepard $(2,2)$-caloron (\ref{hs N-caloron}), with $\lambda_1=\lambda\cos(\beta/2)$, $\lambda_2=\lambda\sin(\beta/2)$, $t_1=-t_2=-\xi/\mu_0$, and $\mathbf{a_1}=-\mathbf{a_2}=(0,-\chi,0)$.

The Nahm data (\ref{xi neq 0}), (\ref{xi=0}) were first studied in section 3.2 of \cite{vb02}, where a numerical Nahm transform was also implemented.  There, the $(2,2$)-calorons were analysed in terms of their four constituent monopoles.  The consituent monopoles appeared as lumps of action density, localised in $\mathbb{R}^3$ and smeared out over $S^1$.  
These lumps were located on the $x^2$-axis, at the points $x^2=y_p^{\pm}$, where $y_1^-$, $y_1^+$ are the eigenvalues of $Y_1$ and $y_2^-$, $y_2^+$ are the eigenvalues of $Y_2$.  

Interestingly, these eigenvalues satisfy the inequalities,
\begin{equation}
\label{eigenvalue inequality}
y_1^- \leq y_2^- \leq y_1^+ \leq y_2^+.
\end{equation}
A proof of this is as follows.  Let $Y_p'$ denote the traceless part of $Y_p$, and let $\|A\|:=\sqrt{-\det(A)}$ for any traceless hermitian matrix $A$.  Then the eigenvalues of $Y_p$ are 
\[ y_p^{\pm}= \frac{1}{2}\mbox{Tr}(Y_p)\pm\|Y_p'\|, \]
and (\ref{eigenvalue inequality}) is equivalent to
\begin{eqnarray}
\nonumber \frac{1}{2}\mbox{Tr}(Y_2-Y_1) &\geq& \|Y_2'\|-\|Y_1'\| \\
\label{inequality 1} \frac{1}{2}\mbox{Tr}(Y_2-Y_1) &\geq& \|Y_1'\|-\|Y_2'\| \\
\nonumber \frac{1}{2}\mbox{Tr}(Y_2-Y_1) &\leq& \|Y_1'\|+\|Y_2'\|.
\end{eqnarray}
It follows from the matching conditions (\ref{explicit matching condition}) that $(1/2)\mbox{Tr}(Y_2-Y_1) = \|Z_2-Z_1\|$, where
\begin{eqnarray*}
Z_1&=&\exp(\xi\nu_1\sigma^3/2\mu_0) Y_1' \exp(-\xi\nu_1\sigma^3/2\mu_0) \\
Z_2&=&\exp(-\xi\nu_2\sigma^3/2\mu_0) Y_2' \exp(\xi\nu_2\sigma^3/2\mu_0).
\end{eqnarray*}
It is also true that $\|Z_p\|=\|Y_p'\|$, and this can be substituted to the right hand sides of the inequalities (\ref{explicit matching condition}).  Thus (\ref{inequality 1}) can be shown to follow from the triangle inequality for $\|\cdot\|$.

\subsubsection{$R_N$ non-trivial}
\label{RN non-trivial}

When $k=\pm1$ the representation $R_N$ is not trivial.  We will only consider $k=1$ here ($k=-1$ is similar up to minus signs); the invariant Nahm matrices must take the form
\begin{eqnarray*}
T_p^0(s) &=& h_p(s)\sigma^2 + b_p(s) \mathbf{1}_2 \\
T_p^1(s) &=& f_p(s)\exp(i\theta_p(s)\sigma^2) \sigma^1 \\
T_p^2(s) &=& g_p(s)\sigma^2 + a_p(s)\mathbf{1}_2 \\
T_p^3(s) &=& f_p(s)\exp(i\theta_p(s)\sigma^2) \sigma^3
\end{eqnarray*}
for $p=1,2$, where $f_p,g_p,h_p,\theta_p,a_p,b_p$ are real functions of $s\in I_p$.  By gauge transformations, we can make $T_p^0=0$.  As we shall see, the Nahm equations and matching conditions can be solved when $l=\pm2$, but the cannot be solved when $l=0$.  When $l=\pm2$ the invariant forms for $W$ and $\tau$ are
\begin{eqnarray*}
W &=& \begin{cases}
       \lambda \exp(\alpha e_2) (e_0,e_2) & l=2 \\
       \lambda \exp(\alpha e_2) (e_1,e_3) & l=-2 \\
      \end{cases} \\
\tau &=& e_2
\end{eqnarray*}
where $\lambda>0$ and $\alpha\in[0,2\pi)$.  One can make $\alpha=0$ by gauge transformation.

The Nahm equations for this invariant Nahm data are
\begin{eqnarray*}
f_p'+2f_pg_p &=& 0 \\
f_p\theta_p' &=& 0 \\
g_p'+2f_p^2 &=& 0 \\
a_p' &=& 0,
\end{eqnarray*}
and the matching conditions are
\begin{eqnarray*}
f_2(\nu_1/2)\cos(\theta_2(\nu_1/2)) - f_1(\nu_1/2)\cos(\theta_1(\nu_1/2)) &=& 0 \\
f_2(\nu_1/2)\sin(\theta_2(\nu_1/2)) - f_1(\nu_1/2)\sin(\theta_1(\nu_1/2)) &=& 0 \\
g_2(\nu_1/2) - g_1(\nu_1/2) &=& \lambda^2/2 \\
a_2(\nu_1/2) - a_1(\nu_1/2) &=& \pm\lambda^2/2,
\end{eqnarray*}
where $\pm$ corresponds to the choice $l=\pm2$.  The Nahm equations have a trivial solution where $f_p=0$, and $g_p$ is constant.  Since $g_p$ is odd, we must have $g_1=0$, $g_2=0$.  It then follows from the matching conditions that $\lambda=0$, so there are no calorons in this case.

When $f_p$ is non-zero, $\theta_p$ and $a_p$ must both be constant.  The most general solution to the Nahm equations for even $f_p$ and odd $g_p$ are
\begin{eqnarray*}
f_1(s) &=& (D_1/2) \sec (D_1 s) \\
f_2(s) &=& (D_2/2) \sec (D_2 (s-\mu_0/2)) \\
g_1(s) &=& -(D_1/2) \tan (D_1s) \\
g_2(s) &=& -(D_2/2) \tan (D_2(s-\mu_0/2))
\end{eqnarray*}
for real constants $D_1$ and $D_2$, which can be assumed to be positive without loss of generality.  In order that these functions remain finite, we must have $D_1 < \pi/\nu_1$ and $D_2 < \pi/\nu_2$.  The matching conditions are easily seen to reduce to,
\begin{eqnarray}
\label{theta mc}
\theta_2 &=& \theta_1 \\
\label{D mc}
D_2 \sec(D_2\nu_2/2) &=& D_1 \sec(D_1\nu_1/2) \\
\label{a mc}
a_2 - a_1 &=& \pm\lambda^2/2 \\
\label{lambda mc}
D_1\tan(D_1\nu_1/2) + D_2 \tan(D_2\nu_2/2) &=& \lambda^2.
\end{eqnarray}

Solving the matching conditions amounts to solving (\ref{D mc}).  This equation admits a one-parameter family of solutions, since the functions $D_p \mapsto D_p\sec(D_p\nu_p/2)$ are bijections from the intervals $(0,\pi/\nu_1)$ to $(0,\infty)$, hence invertible.  However, an explicit parametrisation of the solutions is not known.  Given a solution $(D_1,D_2)$ of (\ref{D mc}), the constants $\lambda$, $a_1$, $a_2$ are determined by (\ref{lambda mc}) and (\ref{a mc}).  Notice that $a_1$ and $a_2$ are only determined up to addition of a constant, and that this constant can be fixed by $x^2$-translations.  Notice also that $\theta_1$ (and hence $\theta_2$) can be made zero by gauge rotation.  Therefore, this family of calorons depends on one parameter, if the position is fixed.

These expressions remain valid in the case where one monopole is massless ($\nu_2=0$); such calorons were considered in \cite{ward}.

In the case where $\nu_1=\nu_2$, the family of calorons considered here forms a subset of the ``rectangular'' configurations of Bruckman et al.  Specifically, in section 6.3 of \cite{vb03} an exact analytic solution of the Nahm equations with $\nu_1=\nu_2$ is given in terms Jacobi elliptic functions.  Our calorons correspond to the case where the elliptic parameter is zero.  Bruckman et al. did not consider the case where $\nu_1\neq\nu_2$.  In reference \cite{vb04} the action densities of these calorons were constructed, making use of a numerical implementation of the Nahm transform.  The action density is concentrated in two rings in $\mathbb{R}^3$, smeared out over $S^1$.  The rings are centred on the $x^2$-axis and lie in planes perpendicular to the $x^2$-axis.  The location of each ring is $x^2=-a_p$.  The rings are identified with the two constituent 2-monopoles.

So far we have only considered Nahm data for which $l=\pm2$.  We also need to consider the case $l=0$; it will turn out that the Nahm equations cannot be solved in this case.  $W$ takes the form
\[ W = \lambda q (1,-e_2)  \]
where $\lambda>0$ and $q$ is a quaternion of unit length ($q^\dagger q=1$).  Since $R_2'$ is trivial, there is no restriction on $\tau$.  The solution of the Nahm equations proceeds in much the same way as before, but the matching condition (\ref{lambda mc}) is replaced by
\[ - D_2\tan(D_2/\nu_2) - D_1\tan(D_1\nu_1/2) = \lambda^2. \]
This can not be solved for nonzero $D_1$ and $D_2$, because the left hand side is less than or equal to zero.  Therefore no calorons are obtained in this case.

\subsection{$U(1)$-symmetric $(2,1)$-calorons}
\label{(2,1)-calorons}

Now we will look for $U(1)$-symmetric $(2,1)$-calorons.  We start by considering the representations acting on the Nahm data in (\ref{T1j sym})-(\ref{T2j sym}).  We set $R_N=Q_k(R)$; then representation theory calculations tell us that we only need to consider invariant Nahm data in the cases $k=0,\pm1$.

When $k=0$ (and $R_N$ is trivial), the invariant Nahm data must have $T_p^1=0$ and $T_p^3=0$ for $p=1,2$.  We choose a gauge so that $T_2^0=0$ and
\begin{eqnarray*}
T_1^0 &=& (\xi/\nu_1)\sigma^3 \\
X &=& \left( \begin{array}{c} \sin(\beta/2) \\ -\cos(\beta/2) \end{array} \right)
\end{eqnarray*}
for $\xi\in[0,\pi]$ and $\beta\in[0,2\pi)$.  We write $T_1^2$ in the form,
\[ T_1^2(s) = \exp( iT_1^0 s) Y \exp(-iT_1^0 s), \]
where $Y$ is a Hermitian $2\times2$ matrix.  The Nahm equations imply that $Y$ and $T_2^2$ are constant.  We write $Y=Y^0\mathbf{1}_2+Y^j\sigma^j$, and the reality condition (\ref{reality condition}) implies that $Y^2=0$.  The matching conditions (\ref{matching condition 3}), (\ref{matching condition 4}) for this Nahm data are,
\[ T_2^2 = Y^0 - Y^3\cos\beta - Y^1\sin\beta\cos\xi. \]
We choose to fix $Y^0=0$ by making an $x^2$-translation.  Then we are left with a four-parameter family of calorons, parametrised by $\xi$, $\beta$, $Y^1$ and $Y^3$.  Note that when $\xi=0$ the gauge is no longer fixed, so $\xi=0$ corresponds to a two-parameter subfamily.  This Nahm data has not been considered before in the literature.  We expect that there is some overlap between this family of calorons and those studied by Chakrabarti in \cite{chakrabarti87}.

These expressions remain valid in the case where the 1-monopole becomes massless, $\nu_2=0$.  In particular, when $\nu_2=0$ and $Y^1=0$, the calorons obtained from this Nahm data are JNR $(2,1)$-calorons (\ref{jnr N-caloron}), with $t_1=-t_2=-\xi/\mu_0$, $\mathbf{a}_1=-\mathbf{a}_2=(0,-Y^3,0)$, $\lambda_1=\cos(\beta/2)$ and $\lambda_2=\sin(\beta/2)$.  On the other hand, when the 2-monopole becomes massless, the caloron is just a 1-monopole.  This is because, when $\mu=0$, a caloron with $k=1$ can be gauge rotated to a caloron with $k=-1$.

By analogy with $(2,2)$-calorons, we expect the constituent monopoles of these $(2,1)$-calorons to be located on the $x^2$-axis with $x^2$-coordinates given by the eigenvalues of the matrices $T_1^2$ and $T_2^2$.  If we denote the eigenvalues of $T_1^2$ by $y_1^-\leq y_1^+$, and write $y_2=T_2^2$, then one can show that
\[ y_1^- \leq y_2 \leq y_1^+. \]
This should be compared with the constraint (\ref{eigenvalue inequality}) on the monopole locations for the family of $(2,2)$ calorons with $R_N$ trivial.

We should also consider invariant Nahm data with $k=\pm1$.  However, no new calorons are obtained this way; the only solutions to the Nahm equations correspond to a subset of the Nahm data found for $k=0$.

\section{Large scale limits}
\label{large scale limits}

The large scale limit of a $(1,1)$-caloron is a 1-monopole \cite{rossi,ll}.  For higher charges, the large scale limit of a caloron may be a monopole \cite{chakrabarti87}, or it may be a caloron with constituent monopoles of unequal charge (as was the case for the Harrington-Shepard calorons).  In this section we will take large scale limits of the $U(1)$-symmetric calorons constructed above.

As was observed in \cite{ll} for charge 1, taking infinite scale limits is the same as sending monopole locations to infinity.  There are  a limited number of ways of doing this for $(2,2)$-calorons.  Either we send a 2-monopole to infinity, leaving behind a $(2,0)$-caloron (that is, a 2-monopole), or we send a 1-monopole to infinity, leaving a $(2,1)$-caloron.  Since for the $U(1)$ symmetric 2-monopoles, the representation $R_N$ on the Nahm data is non-trivial, one would expect the large scale limit of a $(2,2)$-caloron with non-trivial $R_N$ to be a 2-monopole.  One would anticipate that $(2,2)$-calorons with $R_N$ trivial have a $(2,1)$-caloron as their large scale limit, for similar reasons.  We will see that these predictions turn out to be correct; we will also show that the large scale limit of a symmetric $(2,1)$-caloron is a $(1,1)$-caloron.  We note that similar limits have been considered previously for monopoles \cite{chen&weinberg}.

\subsection{$(2,2)$-calorons with $R_N$ non-trivial}

First we consider the family of $(2,2)$-calorons with $R_N$ non-trivial, described in section \ref{RN non-trivial}.  The large scale limit will be $\lambda\rightarrow\infty$.  From equation (\ref{a mc}) we see that $a_2-a_1\rightarrow\pm\infty$ in this limit.  Recalling the interpretation of $a_1$ and $a_2$ as 2-monopole locations, this means that the separation of the monopoles tends to infinity.  We will fix the first monopole at $a_1=0$ so that $a_2\rightarrow\pm\infty$, as illustrated in figure 1.

From equation (\ref{lambda mc}), $\lambda\rightarrow\infty$ implies that $D_1\rightarrow\pi/\nu_1$ or $D_2\rightarrow\pi/\nu_2$.  But from equation (\ref{D mc}), $D_1\rightarrow\pi/\nu_1$ if and only if $D_2\rightarrow\pi/\nu_2$.  We conclude that, in the limit where $\lambda\rightarrow\infty$, the Nahm matrices on the interval $I_2$ diverge, while the Nahm matrices on the interval $I_1=(-\mu,\mu)$ must converge to
\begin{eqnarray*}
T_1^2 &\rightarrow& -(\pi/\mu)\tan(\pi s/2\mu)\sigma^2 \\
T_1^j &\rightarrow& (\pi/\mu) \sec(\pi s/2\mu)\sigma^j \mbox{ for }j=1,2.
\end{eqnarray*}
This is just the Nahm data for a 2-monopole with mass $\nu_1=2\mu$.

\begin{figure}[t]
\psfig{file=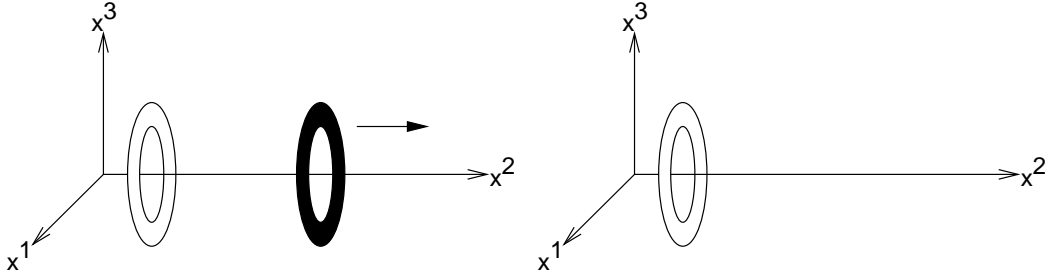, scale=1}
\caption{The large scale limit of a $(2,2)$-caloron with $R_N$ non-trivial.  Each ring represents a constituent 2-monopole.}
\end{figure}

\subsection{$(2,2)$-calorons with $R_N$ trivial}

Now we consider the family of $(2,2)$-calorons with $R_N$ trivial, described in section \ref{RN trivial}.  Recall that the eigenvalues $y_1^\pm$ and $y_2^\pm$ of $Y_1$ and $Y_2$ were interpreted as the four constituent monopole locations.  Our large scale limit will send only one of these four eigenvalues to infinity, as shown in figure 2.

We consider the case $l=1$.  In terms of the parametrisation given in (\ref{xi neq 0}), the eigenvalues are
\begin{eqnarray*}
y_1^\pm &=& \pm \sqrt{ \chi^2 + \frac{\lambda^4}{16} \frac{\sin^2(\beta) \sin^2(\alpha-\xi\nu_2/\mu_0)}{\sin^2\xi} } \\
y_2^\pm &=& \frac{\lambda^2 }{4} \pm \sqrt{ \chi^2 + \chi \frac{\lambda^2}{2} \cos\beta + \frac{\lambda^4}{16} \left( \cos^2\beta + \frac{\sin^2(\beta) \sin^2(\alpha+\xi\nu_1/\mu_0)}{\sin^2\xi} \right) }.
\end{eqnarray*}
$y_1^\pm$ will stay finite in the limit where $\lambda\rightarrow\infty$ and $\alpha\rightarrow \xi\nu_2/\mu_0$, such that
\[ \eta := \lim_{\lambda\rightarrow\infty,\,\alpha\rightarrow \xi\nu_2/\mu_0} \frac{\lambda^2}{4} \frac{\sin(\beta) \sin(\alpha-\xi\nu_2/\mu_0)}{\sin\xi}  \]
is finite.  In this limit, we have
\begin{eqnarray*}
Y_1 & \rightarrow & \chi\sigma^3+\eta\sigma^1 \\
y_2^+ & \rightarrow & \infty \\
y_2^- & \rightarrow & -\eta\cos\xi\sin\beta - \chi\cos\beta \\
v^- &\rightarrow & \left( \begin{array}{c} \sin(\beta/2) \\ -\cos(\beta/2) \end{array} \right),
\end{eqnarray*}
where $v^-$ is the eigenvector of $Y_2$ with eigenvalue $y_2^-$.

In order to get sensible Nahm data in this limit, we need to choose a gauge where $T_2^2$ is constant and diagonal.  We make the gauge transformation $g(s)=g_2g_1(s)$, where
\begin{eqnarray*}
g_1(s) &=& \begin{cases} \exp \left( i \frac{\xi}{\mu_0}\frac{\nu_2}{\nu_1} s \sigma^3 \right) & s\in I_1 
\\ \exp \left( -i\frac{\xi}{\mu_0} \left(s-\frac{\mu_0}{2} \right) \sigma^3 \right) & s\in I_2 \end{cases}
\end{eqnarray*}
makes $T_2^0=0$ and $g_2$ is the matrix which diagonalises $Y_2$.  Now we take the large scale limit, discarding the part of $Y_2$ which becomes infinite, and undo the gauge rotation $g_2$.  We are left with Nahm data,
\begin{eqnarray*}
T_1^0(s) &=& (\xi/\nu_1)\sigma^3 \\
T_1^2(s) &=& \exp(iT_1^0s)(\chi\sigma^3+\eta\sigma^1)\exp(-iT_1^0s) \\
T_2^2(s) &=& - \eta\cos\xi\sin\beta - \chi\cos\beta.
\end{eqnarray*}
This should be compared with the $(2,1)$ Nahm data found in section \ref{(2,1)-calorons}.  We see that this Nahm data is exactly what we had before, with $Y^1=\eta$ and $Y^3=\chi$ and $X=\lim v^-=(\sin(\beta/2),-\cos(\beta/2))^t$.

\begin{figure}[t]
\psfig{file=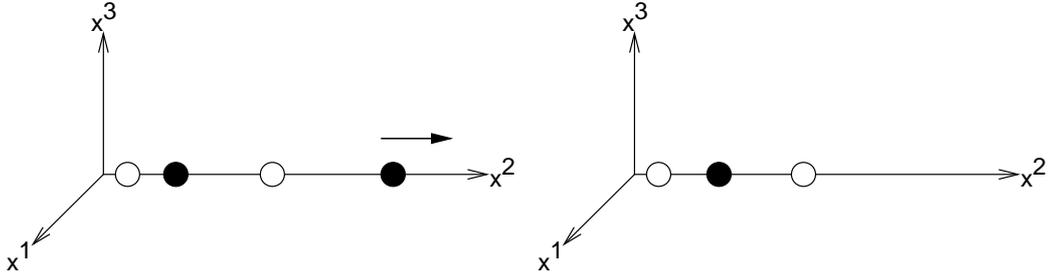, scale=1}
\caption{The large scale limit of a $(2,2)$-caloron with $R_N$ trivial.  The blobs represent constituent monopoles; the white blobs have locations $x^2=y_1^\pm$ and the black blobs have locations $x^2=y_2^\pm$.}
\end{figure}

\subsection{$(2,1)$-calorons}

Finally, we show that the large the large scale limits of the $(2,1)$-calorons described in section \ref{(2,1)-calorons} are $(1,1)$-calorons.  We begin with the eigenvalues of $T_1^2$ and $T_2^2$:
\begin{eqnarray*}
y_1^{\pm} &=& Y^0\pm\sqrt{(Y^1)^2+(Y^3)^2} \\
y_2 &=& Y^0 - Y^3\cos\beta - Y^1\sin\beta\cos\xi.
\end{eqnarray*}
Recall that these are interpreted as constituent monopole locations.  We can take a limit where $Y^0\rightarrow\infty$, $Y^3\rightarrow\infty$, $Y^1\rightarrow0$ and $\beta\rightarrow\pi$, such that $y_1^+\rightarrow\infty$ and $y_2$ and $y_1^-$ stay finite.  This is the large scale limit of the $(2,1)$-caloron.  To obtain sensible Nahm data in this limit, one should choose a gauge so that $T_1^2$ is constant and diagonal (which can always be done).  By taking the large scale limit and discarding the part of this matrix which tends to infinity, we are left with $1\times1$ Nahm matrices $T_1^2=\lim y_1^-$, $T_2^2=\lim y_2$.  These are Nahm matrices for a $(1,1)$-caloron, and the Nahm data is completed by taking $W=\sqrt{2} \lim(y_2-y_1^-)$.

\section{Large period limits}
\label{large period limits}

In this section we will consider large period limits of calorons.  One can show that the $\mu_0\rightarrow0$ limit of caloron Nahm data is ADHM data for an instanton, so the large period limit of a caloron is an instanton.  We will study the ADHM data for the large period limit of the $(2,2)$-calorons constructed earlier; we will see that two distinct families of $U(1)$-symmetric 2-instanton are obtained.  Previous, large period limits of calorons had been considered only in the case where one constituent monopole is massless \cite{ward}.

First we consider Nahm data for an $(N,N)$-caloron.  Let $\nu_1', \nu_2'\in[0,1]$ such that $\nu_1'+\nu_2'=1$.  Suppose that $\epsilon>0$ and that for each $\mu_0\in(0,\epsilon)$, $(T_1,T_2,W)(\mu_0)$ are a set of Nahm data which solve the Nahm equation and matching conditions with $\nu_p=\mu_0\nu_p'$.  We will assume that for each $\mu_0\in(0,\epsilon)$, $T_1$ and $T_2$ have Taylor expansions which converge $T_1,T_2$ on the intervals $I_1,I_2$.

By considering the Nahm equation (\ref{nahm equation}) at $s=0,\mu_0/2$, and substituting the Taylor expansions for $T_p$ into the matching conditions (\ref{matching condition 1}), (\ref{matching condition 2}), we arrive at the following four equations, true for all $\mu_0\in(0,\epsilon)$:
\begin{eqnarray*}
\Im \left( T_1'(0) \right) &=& -i\Im (T_1(0)T_1(0)^\dagger) \\
\Im \left( T_2'(\mu_0/2) \right) &=& -i\Im(T_2(\mu_0/2)T_2(\mu_0/2)^\dagger) \\
2\Im\left( T_2(\mu_0/2)-T_1(0) \right) + O(\mu_0^2) &=& -i\Im(W^\dagger\tau W) \\
-\Im\left( \nu_1' T_1'(0)+\nu_2'T_2'(\mu_0/2) \right) + O(\mu_0) &=& -i\Im(W^\dagger W)/\mu_0.
\end{eqnarray*}
In addition, we suppose that the gauge has been chosen so that $\Re(T_1)$ and $\Re(T_2)$ are constant on the intervals $I_p$ and equal, as in section \ref{nahm transform}.

It follows from the above that if the matrix
\[ \Delta := \lim_{\mu_0\rightarrow0} \left( \begin{array}{c} W/\sqrt{\mu_0} \\ T_1(0)^\dagger \end{array} \right) \]
exists, then it satisfies the ADHM equation, $\Im (\Delta^\dagger \Delta) = 0$.

In a similar way, if $(U_1,U_2,V)$ are solutions of (\ref{dirac equation})-(\ref{dirac matching 2}) for the Nahm data $(T_1,T_2,W)(\mu_0)$ which are well-approximated by their Taylor expansions, then
\[ u(x) := \lim_{\mu_0\rightarrow0} \left( \begin{array}{c} V(x) \\ U_1(0,x)/\sqrt{\mu_0} \end{array} \right) \]
solves $(\Delta+x)^\dagger u=0$ (if the limit exists).  Also (\ref{normalisation}) implies that $u^\dagger u=e_0$ and (\ref{gauge field}) implies that $\lim_{\mu_0\rightarrow0} A_\alpha = u^\dagger \partial_\alpha u$.  Therefore, the whole of the ADHM construction is recovered in a large period limit of the $(N,N)$ Nahm construction.

Now we consider the large period limit of the $(2,2)$-calorons with $R_N$ trivial (section \ref{RN trivial}).  We let $\xi':=\xi/\mu_0$ and $\lambda':=\lambda/\sqrt{\mu_0}$ stay fixed as $\mu_0\rightarrow0$.  Then the ADHM matrix is
\[ \Delta = \left( \begin{array}{cc} \lambda'\cos(\beta/2)e_0 & \lambda'\sin(\beta/2)\exp(\alpha e_2) \\ 
                                     \xi' e_0-\chi e_2 & -\lambda'^2/(4\xi')\sin\beta\sin\alpha e_2 \\
                                     -\lambda'^2/(4\xi')\sin\beta\sin\alpha e_2 & -\xi'e_0 + \chi e_2 
                   \end{array} \right) \]
when $\xi'\neq0$, and
\[ \Delta = \left( \begin{array}{cc} \lambda'\cos(\beta/2)e_0 & \lambda'\sin(\beta/2)e_0 \\ 
                                     -\chi e_2 & 0 \\ 0 & \chi e_2 \end{array} \right) \]
when $\xi'=0$.

To take the large scale limit of the $(2,2)$-calorons with $R_N$ non-trivial (section \ref{RN non-trivial}), we let $\lambda'=\lambda/\sqrt{\mu_0}$ stay fixed as $\mu_0\rightarrow0$.  In this limit, the matching conditions (\ref{D mc}), (\ref{lambda mc}) are solved by $D_1=D_2=\sqrt{2}\lambda'$.  The ADHM matrix is
\[ \Delta = \frac{\lambda'}{\sqrt{2}} \left( \begin{array}{cc} \sqrt{2} e_0 & \sqrt{2} e_2 \\ -e_3 & -e_1 \\ -e_1 & e_3 \end{array} \right). \]

There is a very simple geometrical description of the 2-instanton moduli space, due to Hartshorne \cite{hartshorne,am}.  Associated to each 2-instanton is a circle and an ellipse in $\mathbb{R}^4$ satisfying the Poncelet condition, which states that there exists a one-parameter family of triangles whose vertices lie on the circle and whose edges are tangent to the ellipse.  There are two ways in which the circle and ellipse can be invariant under $U(1)$ rotations: either they lie in the axis of rotation (the plane $x^1=x^3=0$); or they lie in a plane fixed by the rotation ($x^2=$constant and $x^0=$constant), such that the ellipse is a circle, and the centres of both circles sit on the axis of rotation.  It can be shown that the instanton limits of the calorons with $R_N$ trivial are of the first type, while the instanton limits of the calorons with $R_N$ non-trivial are of the second type.

Finally, one can take the large period limits of $(N,N-1)$-calorons.  The large period limit of an $(N,N-1)$-caloron is an $(N-1)$-instanton; in particular, the large period limits of the $U(1)$-symmetric $(2,1)$-calorons are 1-instantons.

\section{Conclusion}
\label{conclusion}

The goal of this article has been to investigate the ways in which calorons can be related to monopoles and instantons.  We found all examples of rotationally symmetric $(2,2)$- and $(2,1)$-calorons by using a combination of the Nahm transform and representation theory.  For all our examples of calorons, we were able to take large period and large scale limits.  The large period limits were instantons, as we expected.  The large scale limits could be either monopoles, or calorons with different magnetic charges.  We found that the representation theory distinguished between these cases.

The constituent monopole interpretation of calorons was central to the taking of large scale limits.  We defined the large scale limit to be the limit where one or more monopole location goes to infinity.  So in general, the large scale limit of a $(m_1,m_2)$-caloron is a $(m_1',m_2')$-caloron, where $m_1'\leq m_1$ and $m_2'\leq m_2$.  It would be interesting to see further examples of calorons with large scale limits.  One way to find more examples of calorons would be to look for Nahm data invariant under discrete symmetry groups; this has already been done in the case of $(N,N)$-calorons with monopole masses $(\nu_1,0)$ \cite{ward}.

Another important question is that of the dimension of moduli space of calorons.  The constituent monopole interpretation would suggest that the moduli space of suitably framed $(m_1,m_2)$-calorons has dimension $4(m_1+m_2)$, since each constituent monopole should contribute three parameters associated with its location and one parameter associated with its $U(1)$ phase.  This prediction turned out to be correct in the case $(m_1,m_2)=(N,N)$ \cite{nogradi,etesi}, but the moduli spaces of calorons with $m_1\neq m_2$ have not yet been studied in any detail.

The constituent monopole picture makes predictions for the number of parameters for the families of $(2,2)$- and $(2,1)$-calorons with $R_N$ trivial.  Since both of these families consist of monopoles on the $x^2$-axis, we would anticipate they depend on repectively eight and six parameters in total, when suitably framed.  The family of $(2,2)$ calorons does depend on eight parameters, including three parameters associated with $x^0$ and $x^2$ translations and a global $U(1)$ gauge rotation.  The unframed family of $(2,1)$ calorons depends on six parameters.  We anticipate that there exists a process for framing $(m_1,m_2)$-calorons with $m_1\neq m_2$, such that the framed moduli space has dimension $4(m_1+m_2)$, and that the family of framed $U(1)$-symmetric $(2,1)$-calorons depends on six parameters.

\subsubsection*{Acknowledgements}
My work is supported by PPARC.  I would like to thank Prof. R. S. Ward for many useful discussions.

\bibliographystyle{hplain}
\bibliography{symcal}

\end{document}